\documentclass[prl,nofootinbib,twocolumn,superscriptaddress]{revtex4}
\def\mysection#1{{\bf #1.} }
\def\mysections#1{{\bf #1.} }

\newcommand{\be}{\begin{equation}}
\newcommand{\ee}{\end{equation}}
\newcommand{\bea}{\begin{eqnarray}}
\newcommand{\eea}{\end{eqnarray}}
\newcommand{\beq}{\begin{equation}}
\newcommand{\eeq}{\end{equation}}
\def\beqa{\begin{eqnarray}}
\def\eeqa{\end{eqnarray}}
\newcommand{\no}{\nonumber}
\def\lsim{\mathrel{\rlap{\lower4pt\hbox{\hskip1pt$\sim$}}
    \raise1pt\hbox{$<$}}}         
\def\gsim{\mathrel{\rlap{\lower4pt\hbox{\hskip1pt$\sim$}}
    \raise1pt\hbox{$>$}}}         
\begin{document}

{\hspace*{13cm}\vbox{\hbox{WIS/36/03-Dec-DPP}
  \hbox{hep-ph/0401012}}}

\vspace*{-10mm}

\title{\boldmath Baryogenesis from the Kobayashi-Maskawa Phase}

\author{Micha Berkooz}\thanks{Incumbent of the Recanati Career Development chair
  for energy research}\email{micha.berkooz@weizmann.ac.il}
\affiliation{Department of Particle Physics,
  Weizmann Institute of Science, Rehovot 76100,
  Israel}

\author{Yosef Nir}\email{yosef.nir@weizmann.ac.il}
\affiliation{Department of Particle Physics,
  Weizmann Institute of Science, Rehovot 76100, Israel}

\author{Tomer Volansky}\email{tomer.volansky@weizmann.ac.il}
\affiliation{Department of Particle Physics,
  Weizmann Institute of Science, Rehovot 76100, Israel}

\vspace*{1cm}

\begin{abstract}
The Standard Model fulfills the three Sakharov conditions for
baryogenesis. The smallness of quark masses suppresses, however, the
CP violation from the Kobayashi-Maskawa phase to a level that is many
orders of magnitude below what is required to explain the observed
baryon asymmetry. We point out that if, as a result of time variation
in the Yukawa couplings, quark masses were large at the time of the
electroweak phase transition, then the Kobayashi-Maskawa mechanism
could be the source of the asymmetry. The Froggatt-Nielsen mechanism
provides a plausible framework where the Yukawa couplings could all be
of order one at that time, and settle to their present values before
nucleosynthesis. The problems related to a strong first order
electroweak phase transition may also be alleviated in this framework.
Our scenario reveals a loophole in the commonly held view that the
Kobayashi-Maskawa mechanism cannot be the dominant source of CP violation
to play a role in baryogenesis.
\end{abstract}

\maketitle

\mysection{Introduction}
The Standard Model (SM) fulfills all three of Sakharov conditions
\cite{Sakharov:dj} for generating a baryon asymmetry in the
Universe. The model fails, however, to explain the observed value of
the asymmetry, $n_B/s\sim10^{-10}$, for two reasons:
\begin{enumerate}
\item CP violation from the Kobayashi-Maskawa (KM) mechanism
  \cite{Kobayashi:fv} is highly suppressed. Explicitly, the relevant
  suppression factor \cite{Jarlskog:1985ht} is given by
  \beq\label{jarmea}
  \epsilon_{\rm CP}=\frac{1}{T_c^{12}}\prod_{i>j \atop u,c,t}(m_i^2-m_j^2)\prod_{i>j \atop
      d,s,b}(m_i^2-m_j^2)J_{\rm CP}\sim10^{-19},
  \eeq
  where $T_c\simeq100$ GeV is the critical temperature at which the electroweak
  phase transition (EWPT) takes place, and $J_{\rm CP}$ is a combination of CKM
  parameters, $J_{\rm CP}\sim s_{12}s_{23}s_{13}\sin\delta_{\rm KM}$.
  \item The EWPT is not strongly first
    order. The experimental lower bound on the Higgs mass, $m_H>114$
    GeV, implies that the EWPT is second order. Consequently,
    sphaleron-induced $(B+L)$-violating interactions are not
    sufficiently suppressed in the broken phase and wash out the
    baryon asymmetry.
  \end{enumerate}
Both failures are related to the numerical values of the Standard
Model parameters. Had quark masses been heavier, the suppression factor
of eq. (\ref{jarmea}) would be milder. Had the Higgs mass been
lighter, the EWPT could be first order. Therefore, if quark and Higgs
masses have been subject to time variations, it is possible that the
Standard Model baryogenesis is the only source of the observed baryon
asymmetry. In this paper, we explore this idea, with focus on four
main issues:
\begin{enumerate}
  \item We are assuming that the Yukawa couplings have been
    time-varying. Is it plausible that the variation was such that
    they all had been of order one in early times?
    \item Cosmology implies that the Yukawa couplings had values close
      to their present ones at the time of nucleosynthesis (NS). Is it
      plausible that the required time variation took place
      between EWPT and NS?
      \item Can we estimate the baryon asymmetry that is produced at
        the EWPT by the KM mechanism with all quark masses of ${\cal
          O}(m_W)$?
      \item What are the implications of time varying couplings on the
        strength of the EWPT?
        \end{enumerate}

\mysection{The Froggatt-Nielsen Mechanism}
The smallness and the hierarchy in the Yukawa couplings are unexplained
within the SM. One of the more attractive frameworks where these
features find a natural explanation is the Froggatt-Nielsen (FN)
mechanism \cite{Froggatt:1978nt}. In this framework, one invokes a
horizontal Abelian symmetry. Fields of
different generations carry different charges under the symmetry. The
symmetry is spontaneously broken by a VEV of a SM-singlet scalar field
$S$. The breaking is communicated to the SM fields through
intermediate quarks and leptons in vector-representations of
the SM gauge group, which have large masses at a scale $M$. The ratio
$\lambda\equiv\langle S\rangle /M$ is assumed to be small,
$\lambda\ll1$. Normalizing the horizontal charges by defining the
charge of $S$ to be $-1$, effective Yukawa couplings suppressed by
$\lambda^{|n|}$ are induced for terms that carry charge $n$ under the FN
symmetry.

In the simplest FN models, the horizontal group is $Z_N$ or $U(1)$,
and it is broken by a single scalar field. The value of $\lambda$ is
often taken to be of order $0.2$ (the value of the Cabibbo angle).
A typical set of FN charges for the quark doublet ($Q_i$) and
antiquark singlet ($\bar U_i$ and $\bar D_i$) fields is the following:
\beqa\label{fncha}
Q_3(0),&\ \ Q_2(+2),&\ \ Q_1(+3),\no\\
\bar U_3(0),&\ \ \bar U_2(+1),&\ \ \bar U_1(+4),\no\\
\bar D_3(+2),&\ \ \bar D_2(+2),&\ \ \bar D_1(+3),
\eeqa
leading to the following parametric suppression of the flavor
parameters (with the FN charge of the Higgs field set to zero):
\beqa\label{supfla}
Y_t\sim1,&\ \ Y_c\sim\lambda^3,&\ \ Y_u\sim\lambda^7,\no\\
Y_b\sim\lambda^2,&\ \ Y_s\sim\lambda^4,&\ \ Y_d\sim\lambda^6,\no\\
s_{12}\sim\lambda,&\ \ s_{23}\sim\lambda^2,&\ \ s_{13}\sim\lambda^3.
\eeqa
With the KM phase being of order one, the suppression factor of
eq. (\ref{jarmea}) is $\epsilon_{\rm CP}\sim\lambda^{28}$.

It is not unlikely that the VEVs of scalar fields have been time
varying. Such a variation could happen for various reasons. One
possibility is that at high enough temperatures, $H\sim T^2/m_{\rm
  Pl}\gg m$, where $m$ is the mass of the scalar, the scalar may be
frozen away from the minimum. Another possibility is that the
minimum of the scalar potential has shifted as a result of finite
temperature effects. Let us consider the case where, for one reason or
another, the value of $\langle S\rangle$ has been time varying, with
$\langle S\rangle\sim M$ prior to the EWPT, compared to its present
value of order $0.2M$. Note that we are considering a rather mild
shift, by a factor of order five. Such a modest variation would,
however, change the Yukawa couplings in a drastic way. There will be
no small parameter to suppress them, and they are all of order one. In
particular, at the EWPT
\beq\label{epspt}
\epsilon_{\rm CP}(T\sim T_c)={\cal O}(1).
\eeq

We learn that in the FN framework, a time variation such that at early
times there is neither smallness nor strong hierarchy in the Yukawa
couplings is not a contrived scenario. Since the smallness of
$\epsilon_{\rm CP}$ is a result of its dependence on a very high
power of a mildly small parameter, once that single parameter is order
one, there is no suppression of $\epsilon_{\rm CP}$.

\mysection{Shifting $\langle S\rangle$ between EWPT and NS}
Our scenario requires that the scalar field $S$ is frozen away from
its present minimum until the EWPT, but to assume its present value
before nucleosynthesis. The simplest mechanism that ensures that the
scalar is frozen until the EWPT but starts to oscillate and redshift
shortly afterwards is by giving it a mass that is close to the Hubble
constant at the time of the EWPT, $H_{\rm EWPT}={\cal O}(10^{-15}\
{\rm GeV})$. Such a light scalar poses, however, problems to cosmology
which are known as `the moduli problem':
\begin{itemize}
\item  Light stable scalars should have masses lighter than the Hubble
constant at the time of matter-radiation equality,
$H_{\rm eq}={\cal O}(10^{-37}\ {\rm GeV})$, in order not to dominate
the energy density of the Universe from rather early times.
\item Light unstable scalars should have decay rates faster than
Hubble constant at the time of nucleosynthesis, $H_{\rm ns}={\cal
  O}(10^{-25}\ {\rm GeV})$.
\end{itemize}
Since $m_S\sim10^{-15}$ GeV, it is too heavy to fulfill the first
condition. Moreover, to avoid being too long lived to fulfill the
second condition, $S$ must decay to final photons or neutrinos with an
effective coupling larger than ${\cal O}(10^{-5})$. An explicit
calculation, assuming that nonrenormalizable couplings are suppressed
by powers of $m_{\rm Pl}$, shows that the actual couplings are much
smaller than that.

Various ways to solve the moduli problem have been suggested in the
literature (see {\it e.g.} \cite{Randall:fr}). Here, we present a
model where the problem is circumvented. Other possible solutions will
be explored in \cite{future}.

The usual scenarios assume that the scalar potential does not change
during the cosmological evolution. We present here a simple model
where the potential does change at the EWPT. Thus, it is not the fast
expansion which freezes $S$ but rather the potential itself. This
situation allows for a larger $m_S$ which, in turn, gives a fast
enough decay rate.

Consider the following potential for the scalar $S$
and the SM Higgs field $\phi$:
\beqa\label{scapot}
V(S,\phi)&=&m_W^4f\left(\frac{S^\dagger S}{M^2}\right)
-\mu_\phi^2\phi^\dagger\phi\left[1+g\left(\frac{S^\dagger S}{M^2}\right)\right]\no\\
&+&\lambda_\phi(\phi^\dagger\phi)^2\left[1+h\left(\frac{S^\dagger
      S}{M^2}\right)\right].
\eeqa
Before EWPT, we have $V(S)=m_W^4f$. After EWPT, we have
$V(S)=m_W^4f-\mu_\phi^2\langle\phi\rangle^2(1+g)
+\lambda_\phi\langle\phi\rangle^4(1+h)$.
We learn that the required shift of $\langle S\rangle$ from $M$ to
$\lambda M$ can occur naturally during the EWPT.

For $M\approx m_{\rm Pl}$, we still run into the cosmological moduli
problem. Assuming that the nonrenormalizable $S$ couplings to SM
fields are suppressed by powers of $M$ and that its mass is of order
$m_W^2/M$ (as is the case for the potential (\ref{scapot})), we find
that the leading decay mode is $S\to\gamma\gamma$ giving $\Gamma\sim
m_S^3/M^2$. Requiring that $S$ decays before nucleosynthesis, we
obtain $M\lsim10^{7}$ GeV. This upper bound on $M$ implies that
$m_S\gsim$ MeV.

The mass $m_S\gsim$ MeV and the decay width $\Gamma_S\gsim10^{-25}$
GeV lie outside of the regions excluded by various astrophysical and
cosmological considerations and by direct laboratory searches
\cite{Masso:1995tw}. A lower bound on $M$ arises from the contribution
of $S$-mediated tree diagrams to flavor changing neutral current
processes. Estimating $\Delta m_K/m_K\sim
[(m_dm_s)/(\lambda^2M^2)](f_K^2/m_K^2)$
(and analogous relations for $D$- and $B$-mixing), we obtain
$M\gsim5\times10^5$ GeV, consistent with the nucleosynthesis
constraint.

In spite of the fact that $M\ll m_{\rm Pl}$, $S$ would start rolling
only at the EWPT because earlier it is stabilized by the potential
rather than by the expansion.

One may ask whether it is possible to have a full high energy theory
that induces (\ref{scapot}) in a natural way. We have been able to
construct a supersymmetric model with a horizontal $Z_N$ symmetry (in
the spirit of models for the scalar potential in
ref. \cite{Leurer:1993gy}) where this is the case. A very large $N$
is, however, required in order to make the model consistent with all
our requirements. We are presently exploring other models
\cite{future}.

\mysection{Suppression from CP Violation}
The possibility that the Standard Model accounts for baryogenesis is
intriguing \cite{Farrar:sp}, but in the end the suppression from the
Kobayashi-Maskawa CP violation [$\epsilon_{\rm CP}$ of
eq. (\ref{jarmea})] is too strong for it to make a significant
contribution \cite{Huet:1994jb,Gavela:dt}. (For SM values of the
flavor parameters, this statement holds model independently and, in
particular, independent of the dynamics of the phase transition
\cite{Huet:1994jb}.) The situation is of course
different in our scenario, where $\epsilon_{\rm CP}={\cal
  O}(1)$.  We now explain which of the usual suppression
factors related to CP violation are lifted in our scenario and which
are not. We also explain why the analysis of the SM baryogenesis
\cite{Huet:1994jb} cannot be simply applied to our case. (An analysis
of SM baryogenesis with quark masses of ${\cal O}(T_c)$ is beyond the
scope of this paper \cite{future}.)

During the EWPT, a bubble of the true vacuum ($\langle\phi\rangle = v
\neq 0$) expands, sweeping out space until it becomes our observed
universe. While the bubble expands, equal number of quarks and
antiquarks hit the bubble wall, some reflected and some
transmitted through the wall. CP violating interactions induce an
asymmetric distribution of quarks and antiquarks. In a simplified
picture where baryon-number violating interactions are infinitely fast
in the unbroken phase, but do not exist in the broken phase, an
excess of baryon number outside the bubble is immediately washed out
while an opposite amount of baryon number is preserved inside. One can
then calculate $n_B/s$.  Taking the thin wall approximation,
ref. \cite{Huet:1994jb} obtains (to leading order in the wall velocity
$v_W$)
\beq\label{hsnbs}
  \frac{n_B}{s} \simeq \frac{10^{-3}}{T^2}\int
  \frac{d\omega}{2\pi}n_0(\omega)(1-n_0(\omega))\Delta(\omega)
  (\vec{p}_L-\vec{p}_R) \cdot \hat v_W.
\eeq
Here $n_0(\omega) = 1/(e^{\omega/T}+1)$ is the Fermi-Dirac
distribution, and $\vec p_{L,R}$ are the left-handed/right-handed quark
momenta. The reflection asymmetry $\Delta(\omega)$ is given by
${\rm Tr}(\bar R^\dagger_{LR}\bar R_{LR} - R^\dagger_{LR}R_{LR})$
where $R_{LR}$($\bar R_{LR}$) is the reflection coefficient for
$q_L\to q_R$ ($\bar q_R\to\bar q_L$).

In the standard scenario, the main suppression factor of $n_B/s$ comes
from $\Delta(\omega)\sim\epsilon_{\rm CP}$. This is expected on
general grounds, since the KM mechanism of CP violation is operative
if and only if $\epsilon_{\rm CP}\neq0$. For Yukawa couplings of order
one, however, this suppression factor is entirely lifted. Another,
milder suppression comes from $(\vec{p}_L-\vec{p}_R) \cdot \hat
v_W/T\sim \alpha_W$ \cite{Huet:1994jb}. This is, again, expected since
usually the dominant interaction that distinguishes left-handed from
right-handed particles is the weak interaction. Yukawa interactions
with the Higgs also distinguish between left- and right-handed quarks.
In our scenario, with Yukawa couplings of order one, the Yukawa
interaction gives the dominant contribution and lifts this suppression
factor too. Thus, in our scenario, there is no parametric suppression
of $n_B/s$, but there are still numerical suppression factors,
$\frac{1}{T}\int\frac{d\omega}{2\pi}n_0(\omega)
(1-n_0(\omega))f_9(\omega)\sim10^{-3}$ (where $f_9(\omega)$ is defined
in ref. \cite{Huet:1994jb}). Combined with the explicit factor of
$10^{-3}$ in eq. (\ref{hsnbs}), related to $g_*$ (the number of
relativistic degrees of freedom), it leads to a suppression of $n_B/s$
by ${\cal O}(10^{-6})$.

At this stage, we cannot make a more precise statement about the size
of $n_B/s$ in our scenario. The reason is not only that the precise
values of the (order one) Yukawa couplings before EWPT are not known,
but also because the calculation is different from the light quark
case in several important ways.
\begin{enumerate}
\item In order to calculate eq. (\ref{hsnbs}), a picture of
quasiparticles was introduced, where the massless quarks acquire a
complex effective mass due to interactions with the hot plasma. In the
standard picture, there are several mass scales: the light quark
masses ($m_d,m_s$), the typical momentum difference between left- and
right-handed quarks ($\sim6$ GeV) and the critical temperature
($\sim100$ GeV). The clear hierarchy between these scales in the
standard scenario no longer exists, and various approximations based
on the hierarchy are not valid.
\item To calculate $\Delta(\omega)$, an effective Dirac equation is
  employed for the quasiparticles. It takes into account the damping
  rate for the quasiparticles at zero momentum,
  $\gamma\simeq0.15g_s^2T\sim20$ GeV. For particle momenta that are
  not much smaller than the temperature, there may be large
  corrections to this result. Furthermore, the effective Dirac
  equation has been solved by treating the quasiparticle masses as
  perturbations \cite{Huet:1994jb}. The perturbative expansion is
  valid only for $m_q/6\gamma\ll1$. Within the standard scenario, this
  is a good approximation to all but the top quark. In our case, this
  is a questionable approximation for all quarks.
\end{enumerate}

Therefore, our analysis above should be used only to identify the
small parameters which still play a role in suppressing the baryon
asymmetry (and those which do not). It should not be taken for a
reliable estimate of other, apriori order one, factors.

\mysection{The Electroweak Phase Transition}
In order for EW baryogenesis to successfully produce the baryon
asymmetry, it is necessary that the EWPT is first order. The condition
of first order phase transition is commonly quantified by demanding
that baryon number that was produced during the phase transition will
not be diluted by baryon violating interactions inside the bubble.  By
calculating the rate of interactions, one obtains the sphaleron bound
on $\langle\phi\rangle$, the VEV of the SM higgs (for a pedagogical
review, see \cite{Quiros:1999jp}):
\beq\label{sphbou}
  {\langle\phi(T_c)\rangle}/{T_c} \gtrsim 1.3.
\eeq
The VEV is related to the Higgs mass and so the above constraint is
translated to a constraint on the mass of Higgs which, at one loop, reads:
\beq\label{higbou}
m_h \lesssim 42\ {\rm GeV}.
\eeq
Such a low value (and the somewhat weaker bound when higher loops are
taken into account) is experimentally ruled out. Thus, not only CP
violation has to have new sources, but also the EWPT has to be
different from the SM one, {\i.e.} the scalar sector must be extended.

In our scenario, the Higgs mass is, in general, different from its
present value. As can be seen from eq. (\ref{scapot}), the corrections
are of order $m_W^2\left(\frac{S^\dagger S}{M^2}\right)={\cal O}(m_W^2)$.
The change is then significant  and could easily take the Higgs mass
to be low enough at the time of EWPT to make it first order.

\mysection{Conclusions}
The Standard Model fulfills all three Sakharov's conditions that are
necessary for baryogenesis. The failure of the Kobayashi-Maskawa
mechanism to account for the observed baryon asymmetry is related to
the numerical values of the Standard Model parameters. The smallness
of the quark flavor parameters (masses and mixing angles) suppresses
CP violation too strongly, while the mass of the Higgs boson is too
heavy for a first order electroweak phase transition to occur.

The possibility that there have been time variations in the Yukawa
couplings opens up a window for the Kobayashi-Maskawa phase to be
the only source of CP violation and to drive baryogenesis. Within the
Froggatt-Nielsen framework, where the structure of the flavor
parameters is a result of an approximate horizontal symmetry, it is
plausible that the time variation has been such that all flavor
parameters were order one at early times.

The dynamics of electroweak baryogenesis when none of the quarks is
much lighter than the electroweak breaking scale may be quite
different from the case of light quarks and needs to be carefully
investigated.  There are no obvious parametric suppression factors and
the produced baryon number may even be as large as ${\cal O}(10^{-6})$. We
conclude that the observed baryon asymmetry can be explained by the KM
mechanism, provided that there are no strong dilution factors coming
from the dynamics.

The possibility that there has been time variation in the Higgs
potential parameters opens up a window for the EWPT to be
first order with only the single Higgs doublet playing a direct
role.

It is worth emphasizing that our scenario can also be implemented in
the supersymmetric framework. The Froggatt-Nielsen selection rules are
somewhat different \cite{Leurer:1992wg}, but the analysis presented
here is unchanged. The EWPT could be first order even without time
variations in the couplings. While in generic supersymmetric models
there are additional sources of CP violation that affect baryogenesis,
our mechanism is particularly interesting for models of minimal flavor
violation, where the CKM matrix is the only source of flavor and CP
violation.

Our scenario introduces a loophole in the commonly held view that the
Kobayashi-Maskawa mechanism cannot be the only source of CP violation
to play a role in baryogenesis.

\mysections{Acknowledgments} We thank T. Banks, R. Brustein, M.
Dine, S. Davidson, W. Fischler, M. Losada, A. Nelson and M. Peskin for useful
discussions. The work of M.B. is supported by the
Israeli Academy of Science centers of excellence program, by the
Minerva Foundation, and by EEC RTN-2000-001122. M.B. would like to
thank the KITP for its hospitality during the final stages of this
project. The research of Y.N. is supported by the Israel
Science Foundation founded by the Israel Academy of Sciences and
Humanities, by a Grant from the G.I.F., the German--Israeli
Foundation for Scientific Research and Development, by a grant
from the United States-Israel Binational Science Foundation (BSF),
Jerusalem, Israel, by the Minerva foundation, and by EEC RTN
contract HPRN-CT-00292-2002. 


\end{document}